\documentclass[preprint,12p]{revtex4}

\usepackage[brazil, english]{babel}
\usepackage[utf8]{inputenc}
\usepackage{graphicx}
\usepackage{epsfig}
\usepackage{amssymb}
\usepackage{amsmath} 
\usepackage[unicode=true,bookmarks=false,breaklinks=false,pdfborder={0 0 1},colorlinks=true]
 {hyperref}
\hypersetup{
 citecolor=blue,linkcolor=blue,urlcolor=blue}

\graphicspath{%
    {converted_graphics/}
    {/}
}

\begin{document}

\title{Joule-Thomson expansion for quantum corrected AdS-Reissner-N\"ordstrom black holes in Kiselev spacetime}

\author{J. P. Morais Graça$^{1}$}
\email[Eletronic address: ]{jpmorais@gmail.com}

\author{Eduardo Folco Capossoli$^{2}$}
\email[Eletronic address: ]{eduardo\_capossoli@cp2.g12.br}
\author{Henrique Boschi-Filho$^{1}$}
\email[Eletronic address: ]{boschi@if.ufrj.br}  

\author{Iarley P. Lobo$^{3,4}$}
\email[Eletronic address: ]{lobofisica@gmail.com}

\affiliation{
$^1$Instituto de F\'{\i}sica, Universidade Federal do Rio de Janeiro, 21.941-972 - Rio de Janeiro-RJ - Brazil \\ 
 $^2$Departamento de F\'{\i}sica and Mestrado Profissional em Práticas de Educa\c{c}\~{a}o B\'{a}sica (MPPEB),  Col\'egio Pedro II, 20.921-903 - Rio de Janeiro-RJ - Brazil\\
 $^{3}$Department of Chemistry and Physics, Federal University of Para\'iba, \\ Rodovia BR 079 - Km 12, 58397-000 Areia-PB,  Brazil.\\
$^{4}$Physics Department, Federal University of Lavras, Caixa Postal 3037, 37200-000 Lavras-MG, Brazil.
}

\begin{abstract}
In this work we study the inversion temperature associated with the Joule-Thomson expansion from the thermodynamics of  AdS-Reissner-Nördstrom black holes. We include quantum corrections in a cosmological fluid that can describe phantom dark matter or quintessence, both in a Kiselev scenario. The description of such physical systems involve numerical solutions and the results are presented as temperature-pressure plots for various values of the parameters of our model. We find non-zero minimum inversion temperatures as well as non-zero minimum pressures depending on the values of those parameters. Completing our study,  we also find isenthalpic curves associated with black hole fixed mass  processes. 
\end{abstract}

\maketitle


\section{Introduction}

Black hole thermodynamics is one of the most interesting topics in General Relativity. The history of the subject goes back to when Bekenstein proposed that the area of the black hole is proportional to its entropy \cite{Bekenstein:1973ur}, followed by Hawking's discovery that black holes radiate \cite{Hawking:1974sw}. Since then, various thermodynamic properties have been attributed to black holes, such as internal energy, entropy, heat capacity, enthalpy, etc.

A little bit more than twenty years after such studies in Refs. \cite{Bekenstein:1973ur, Hawking:1974sw} the iconic work done by Witten, in Ref. \cite{Witten:1998zw}, by using the new-found AdS/CFT correspondence, as  proposed in Ref. \cite{Maldacena:1997re}, relates the Hawking temperature achieved in a curved high-dimensional spacetime to the temperature of a super conformal Yang-Mills  theory in a flat four-dimensional spacetime. Soon after Witten's work the authors in Refs. \cite{Chamblin:1999tk, Chamblin:1999hg} studied the thermodynamic associated to a charged AdS black holes in the holographic context, and then opening up a multitude of possibilities to connect string and gauge theories through various types of black hole and its thermodynamics. It is worthwhile to mention that within holography the thermodynamics quantities are derived from the holographic renormalization of the on-shell euclidian action or the thermodynamics potentials. 

Within general relativity context, some authors have proposed to associate mechanical pressure to black holes. For doing this, they considered that the cosmological constant is a thermodynamic variable that can be associated with the black hole pressure as shown in Refs. \cite{Kastor:2009wy,Kubiznak:2012wp}. In this way, applying the first law of thermodynamics, it is possible to deduce what would be the thermodynamic volume of a black hole, which is different from its physical volume. The existence of mechanical pressure in black holes allows us a better analogy with usual mechanical systems, such as thermal machines, where phase transitions occur naturally. For example, Johnson, in Ref. \cite{Johnson:2014yja}, calculates the efficiency rate of a black hole, considering it to be a thermal machine in a Carnot cycle. Recently, another interesting example is presented by \"Okc\"u and Aydıner in Ref. \cite{Okcu:2016tgt} where they have studied a process similar to the Joule-Thomson process, in a black hole, and were able to calculate their temperature inversion curves.
Traditionally, the Joule-Thomson effect aims to study the temperature variation when a gas goes through a process where its pressure changes. In this mechanical system, the temperature can increase or decrease, and for some gases a critical value can be found where a transition occurs when the temperature changes from increasing to decreasing, or the other way around. This can be represented mathematically by: 
\begin{equation}
    \label{JTcoef}
    \mu = \left( \frac{\partial T}{\partial P}\right)_{H}\,,
\end{equation}
\noindent which measures the temperature variation by the pressure variation in an isenthalpic process. Besides $\mu$ is Joule-Thomson coefficient and the mentioned inversion temperature is achieved by taking $\mu=0$ in above equation.

Since the work \cite{Okcu:2016tgt}, several authors have proposed to study the Joule-Thomson process for different configurations of AdS black holes, in different gravitational theories \cite{Okcu:2017qgo,Mo:2018rgq,Lan:2018nnp,Mo:2018qkt,Cisterna:2018jqg,Rizwan:2018mpy,Chabab:2018zix,Yekta:2019wmt,Li:2019jcd,Nam:2019zyk, K.:2020rzl, Bi:2020vcg}. In most of these works, the result obtained follows a well-established pattern, which differs from the traditional result of a Van der Waals gas (see, e. g. \cite{Reif}). 

The main purpose of this work is to study the effect of quantum corrections and cosmological fluids, such as quintessence and phantom dark energy, on the Joule-Thomson expansion in an  AdS-Reissner-Nördstrom (AdS-RN) black hole. 

Quantum corrections are relevant in the phenomenology of microscopic black holes. Its effects in a static black hole have been studied by Kazakov and Solodukhin \cite{Kazakov:1993ha}, where the authors considered small deformations in the Schwarzschild metric due to the quantum fluctuations in the gravitational and matter fields. The effect of quantum corrections in black hole thermodynamics and phase transitions were studied in Ref. \cite{Shahjalal:2019pqb, Lobo:2019put}.

On the other hand, the introduction of a cosmological fluid is important because it is present both in the current and in the early universes. In order to incorporate cosmological fluids, we will use the Kiselev metric \cite{Kiselev:2002dx}, where such matter can be described by an equation of state like $P = \, \omega \, \rho$. For instance, when $\omega = -2/3 $, it can be said that the environment around the black hole is a kind of quintessence, and when $\omega < - 1$ the environment is a kind of phantom dark energy \cite{Caldwell:1999ew}.

This paper is divided as follows. In section \ref{Kiselev}, we will present the AdS-Reissner-Nördstrom black hole with quantum corrections, surrounded by a Kiselev spacetime and make a brief study of its properties that are relevant to our study. In section \ref{thermo}, we will calculate the mass of the black hole, its Hawking temperature and  the inversion temperature of the Joule-Thomson expansion. In section \ref{results}, we will perform a numerical study of the inversion temperature and obtain their graphical solutions and the corresponding isenthalpic curves. Finally, in section \ref{conc}, we will present our conclusions.

\section{Quantum corrected charged AdS black hole surrounded by a Kiselev spacetime}\label{Kiselev}

The spacetime metric of the quantum corrected charged AdS black hole surrounded by a cosmological fluid is given by the following spherically symmetric metric,
\begin{equation}
    ds^2 = f(r) dt^2 - f(r)^{-1} dr^2 - r^2 d\Omega^2,
    \label{secIImetric}
\end{equation}
where $d\Omega^2$ is the solid angle, $d\theta^2 + sin(\theta)^2 d\phi^2$, and the horizon function is
\begin{equation}\label{hor}
     f(r) =  - \frac{2 M}{r} + \frac{\sqrt{r^2 - a^2}}{r} + \frac{r^2}{\ell^2} - {\frac{c}{r^{3\omega + 1}}} + \frac{Q^2}{r^2}.
\end{equation}

The parameter $M$ is the black hole mass, $a$ is related to the quantum correction, $\ell$ is the length scale of the asymptotically AdS spacetime, $c$ is a parameter related with the cosmological fluid, and $Q$ is the electric charge of the black  hole. 

Let us start by explaining why we have chosen the metric above and the origin of each term. First, we can highlight that M. Visser recently proposed that the Kiselev black hole can be generalized to a spacetime with $N$ components whose relationship between energy and pressure, individually, is linear as shown in Ref. \cite{Kiselev:2002dx,Visser:2019brz}. That is, the spacetime for the generalized Kiselev black hole can be written as the metric (\ref{secIImetric}) with
\begin{equation}
    f(r) = \left(1 - \sum_{i=i}^{N} \frac{c_i}{r^{3 \omega_i + 1}}\right).
\end{equation}

\noindent
Thus, choosing appropriate values for $\omega_i$ and $c_i$, one can reproduce all the terms of the proposed metric, except the mass term, because it receives a quantum correction. Nothing prevents us, then, from aggregating several terms whose equation of state is simply $P = \omega_i \rho$, although we must note that this relationship is built from an average in the energy-momentum tensor, as explained in the original article by Kiselev \cite{Kiselev:2002dx}.

That said, we still have to explain why we chose the terms above. If we put $\omega = 1/3$ and $c<0$, this term will be equivalent to an electric charge summing up with the last term in Eq. \eqref{hor}. The anti-de Sitter ($\omega = -1$), represented by the term $r^2/\ell^2$ in Eq. \eqref{hor}, is necessary for the definition of a mechanical pressure for the black hole \cite{Kastor:2009wy,Kubiznak:2012wp}. The novelty in this paper is the introduction of a cosmological fluid term in order to mimic quintessence, for instance, $\omega = -2/3, -1/6$, or phantom dark energy ($\omega = - 4/3$). 

The parameter $a$ is related to a deformation of the black hole mass due to quantum corrections and its origin can be found in \cite{Kazakov:1993ha}. It is an independent parameter and for $a=0$ the metric reduces to the AdS-Reissner-Nordström metric surrounded by a cosmic fluid. In principle, it can assume any value as long as it is smaller than the radius of the event horizon, as expected, since it is a small correction to the black hole metric.

Before we study the thermodynamics of the system, let us make a little digression about the limits of the metric and the coordinate system. At the limit where the radius is small, we have 
\begin{eqnarray}
   f \rightarrow 1 - \frac{2 M}{r} - \frac{a^2}{2 r^2} + \frac{Q^2}{r^2}
\end{eqnarray}

\noindent
as long as $3 \omega + 1 < 1$, which means that the black hole has two horizons, except in the extreme case where $4\,{M}^{2}-4\,{Q}^{2}+2\,a \leq 0$. 

At the limit where the radius goes to infinity, we have
\begin{eqnarray}
   f \rightarrow \frac{r^2}{l^2} - \frac{c}{r^{3 \omega + 1}}
\end{eqnarray}

\noindent
and the existence of a cosmological horizon depends on the values for $\omega$ and $c$. In the case where $ \omega = -2/3$, the space will be asymptotically anti-de Sitter. This means that we must take a little care with our numerical analysis, since there is more than one horizon radius.

\section{Black hole thermodynamics and the inversion temperature}\label{thermo}

In this section we will provide a description of the black hole thermodynamics in our model and obtain the corresponding  inversion temperature associated with the Joule-Thomson process. 
In this sense we will consider the mass of the black hole playing the  role of the enthalpy, the area of the event horizon as the entropy and the cosmological constant will be associated with the black hole's pressure  \cite{Kubiznak:2012wp}.

The first law of thermodynamics can encompass variations of the parameters (hair) that define the black hole, which in this case are its area, cosmological constant, electric charge, quintessence parameter and quantum correction parameter (for details, please see \cite{GBKiselev} for the inclusion of the quintessence parameter as a thermodynamics variable, which we are extending by including the possibility of variation of the quantum correction parameter) can be written as 
\begin{equation}
    \label{firstLaw}
    dH = T dS + V dP + \Phi dQ+{\cal C}dc+{\cal A}da\, ,
\end{equation}

\noindent
and, since we are considering  enthalpy and any non-mechanical work as constants,\footnote{The constancy of thermodynamic quantities other than the mechanical work is the framework in which this analysis is being carried out in the literature, and is also the one followed in this paper.}  
\begin{equation}
    \label{isenthalpic}
    0 = T \left( \frac{dS}{dP} \right)_H + V.
\end{equation}

To relate the above expression to the Joule-Thomson coefficient we must use some thermodynamic relations. Since the entropy is a state function, 
\begin{equation}
    dS = \left( \frac{\partial S}{\partial P} \right)_{T} dP +\left( \frac{\partial S}{\partial T} \right)_{P} dT
\end{equation}

\noindent
or, for an isenthalpic process,
\begin{equation}
    \left( \frac{\partial S}{\partial P}  \right)_H = \left( \frac{\partial S}{\partial P} \right)_{T} + \left( \frac{\partial S}{\partial T} \right)_{P} \left( \frac{\partial T}{\partial P} \right)_{H}.
\end{equation}

Replacing this expression into Eq. (\ref{isenthalpic}), one has
\begin{equation}
    0 = \left[ \left( \frac{\partial S}{\partial P} \right)_{T} + \left( \frac{\partial S}{\partial T} \right)_{P} \left( \frac{\partial T}{\partial P} \right)_{H} \right] + V.
\end{equation}

Finally, using the Maxwell relation $(\partial S / \partial P)_T =  - (\partial V / \partial T)_P$ and the definition for specific heat $C_p = T (\partial S / \partial T)_P$, one can substitute the above expression into (\ref{JTcoef}) to find
\begin{equation}
    \mu = \frac{1}{C_p} \left[ T \left( \frac{\partial V}{\partial T} \right)_P - V \right],
\end{equation}

\noindent
and the inversion temperature occurs when $\mu = 0$, i.e.,
\begin{equation}
    T_i = V \left( \frac{\partial T}{\partial V} \right)_P,
\end{equation}

\noindent
where the system changes from heating to cooling or the other way around. 

Let us now proceed to study this process for the black hole. Fist of all, let us compute the black hole mass $M$, which is defined as a function of the location of its horizon $r_+$, trough the largest root of $f(r_+) = 0$, before any cosmological horizon. Then by using Eq. \eqref{hor}, one gets
\begin{equation}\label{mbh}
    f(r_+) = 0 \Rightarrow  M = \frac{1}{2} \left(\sqrt{r_+^2-a^2}-c r_+^{-3 \omega }+\frac{r_+^3}{\ell^2}+\frac{Q^2}{r_+}\right)\,.
\end{equation}
%
%
%

At this point it is worthwhile to comment some thermodynamical aspects of gravitational parameters appearing in the above equation. First note that $M$ interpreted as the enthalpy of the system. Then, the parameter $a$ plays the role of a correction for the enthalpy. On the other side, the parameter $c$ can be associated with a generalized potential, depending on the power $\omega$. In particular, for $\omega= 1/3$, the parameter $c$ becomes a correction to the electric charge $Q$.

In order to compute the entropy of the quantum-corrected Schwarzschild black hole in Kiselev spacetime, one can see in Ref. \cite{Shahjalal:2019pqb} that this entropy is the same as Hawking-Bekenstein entropy, so that
\begin{equation}\label{entrobh}
    S = \frac{A}{4}\Leftrightarrow S= \frac{4 \pi r^2_+}{4} \Leftrightarrow r_+ = \sqrt{\frac{S}{\pi}}\,.
\end{equation}

From the first law \eqref{firstLaw}, the electric potential $\Phi$, the quintessence potential ${\cal C}$ and the quantum correction potential ${\cal A}$ are
\begin{align}
    &\Phi=\left(\frac{\partial M}{\partial Q}\right)_{S,P,c,a}=Q\sqrt{\frac{\pi}{S}}\, ,\\
    &{\cal C}=\left(\frac{\partial M}{\partial c}\right)_{S,P,Q,a}=-\frac{1}{2}\left(\frac{\pi}{S}\right)^{3\omega/2}\, ,\\
    &{\cal A}=\left(\frac{\partial M}{\partial a}\right)_{S,P,Q,c}=-\frac{a}{2}\left(\frac{S}{\pi}-a^2\right)^{-1/2}\, ,
\end{align}
where the regularization of the black hole singularity is seen through the needed condition $S>\pi a^2$ in the last expression. 

Regarding our last thermodynamics variable, the pressure, we will consider it as related with the cosmological constant. Thus, the pressure has the following form \cite{Kubiznak:2012wp}: 
\begin{equation}\label{pressao}
    P =  \frac{3}{8 \pi\ell^2} \,,
\end{equation}

Replacing Eqs. \eqref{entrobh} and \eqref{pressao} in Eq. \eqref{mbh}, one has:
\begin{equation}
    M = \frac{1}{2\sqrt{\pi}} \left(\sqrt{{S}-\pi a^2}-c \pi ^{\frac{3 \omega +1}{2}} S^{-\frac{3 \omega }{2}}
    +\frac{8 P S^{3/2}}{3} +  \pi  Q^2 S^{-1/2}
     \right).
\end{equation}
The Hawking temperature is given by
\begin{equation}\label{HTnew}
    T_H = \left(\frac{\partial M}{\partial S} \right)_{P,Q}\,, 
\end{equation}{}
and then one can write 
\begin{equation}
    T_H= \frac{1}{4\sqrt{\pi}} \left({\frac{1}{\sqrt{S-\pi  a^2}}+8 P \sqrt{S}}+3 \frac{ c \,\omega}{\sqrt{\pi}} \left(\frac{\pi}{S}\right) ^{\frac{3 \omega }{2}+1} -\frac{{\pi }\, Q^2}{S^{3/2}}
    \right)\,.
\end{equation}
Using Eq. \eqref{entrobh}, one can rewrite the Hawking temperature as a function of the horizon radius, so that
\begin{equation}\label{ti1}
    T_H =\frac{1}{{4 \pi }}\left({\frac{1}{\sqrt{r_+^2-a^2}}+3 c \omega  r_+^{-3 \omega -2}+8 \pi  P r_+ -\frac{Q^2}{r_+^3}}\right)\,.
\end{equation}
Now, using the numerical solutions to Eq. \eqref{mbh} written as $r_+=r_+(a,c,\omega,P, Q, M)$, in this equation, one gets the inversion temperature $T=T(a,c,\omega,P, Q, M)$. This result will be used in next section to construct the isenthalpic curves.

From the  thermodynamics first law, Eq. \eqref{firstLaw}, the thermodynamic volume of the black hole can be computed as 
\begin{equation}
    V = \left( \frac{\partial M}{\partial P}\right)_{S,Q} \Rightarrow V= \frac{4}{3} \pi r^3_+ \,,
\end{equation}
which gives us $r_+ = (\frac{3 V}{4 \pi})^{1/3}$. Furthermore the Hawking temperature as a function of the black hole volume reads
\begin{equation}\label{thpv}
   T_H= \frac{1}{2 \pi}{\left[{\left(\frac{6V}{\pi }\right)^{\frac{2}{3}} -4 a^2}\right]^{-\frac{1}{2}}}
   +\frac{3 \,\omega\, c}{4\pi} \left(\frac{3V}{4\pi }\right)^{-\omega - \frac{2}{3}} + P \left(\frac{6V}{\pi }\right)^{\frac{1}{3} } 
   -\frac{Q^2}{3 V}\,.
\end{equation}

In order to get the inversion temperature, one can use its definition (for fixed $H, Q, c$ and $a$), so that 
\begin{equation}
    T_i = V \left( \frac{\partial T}{\partial V}\right)_P.
\end{equation}
By using Eq. \eqref{thpv} and computing the inversion temperature, one has
\begin{equation}\label{Tir+}
    T_i= -\frac{V^{2/3}}{\sqrt[3]{6} \pi ^{5/3} \left(\left(\frac{6V}{\pi }\right)^{2/3} -4 a^2\right)^{3/2}}-\frac{ c\, \omega\left(3\omega +{2}\right)}{4\pi} \left(\frac{3V}{4 \pi }\right)^{-\omega -\frac{2}{3} }    +\sqrt[3]{\frac{6V}{\pi }} \frac{P}{3} +\frac{Q^2}{3 V},
\end{equation}
or, as a function of $r_+$,
\begin{equation}\label{ti2}
    T_i = \frac{2}{3}  P r_+
    -\frac{r_+^2}{12 \pi\left(r_+^2-a^2\right)^{3/2}}- \frac{c\, \omega  (3 \omega +2)}{4\pi r_+^{2+3\omega}}  + \frac{ Q^2}{4\pi  r_+^3}.
\end{equation}
Finally, subtracting Eqs. \eqref{ti2} and \eqref{ti1}, one gets
\begin{eqnarray}\label{dift}
   {\frac{4r_+^2 - 3 a^2}{\left(r_+^2-a^2\right)^{3/2}}+ \frac{3\, c\, \omega  (3 \omega +5)}{r_+^{3 \omega +2}}-\frac{6 Q^2}{r_+^3}}+ 16\pi P r_+ = 0.
 \end{eqnarray}
This equation can be approximated for small values of $a$. 
In particular, for $c=0$, this equation can be solved analytically and implies the minimum inversion temperature ($P=0$), up to second order in $a$: 
\begin{equation}\label{tmin}
T_{min} = \frac{\sqrt {6}}{36 \,\pi\, Q  } \left[\, 1 +\left(\frac {a}{2 \, Q}\right)^2 \, \right] \,.
\end{equation}
This equation shows the influence of the quantum correction on the minimum inversion  temperature. If one takes the limit $a\to 0$ the result for $T_{min}$ presented in Ref. \cite{Okcu:2016tgt} is recovered. 

On the other hand, the above equations imply a non zero pressure even at zero inversion temperature. 
This can be seen looking, for instance, at Eq. \eqref{ti1} and demanding $T_H=0$, so that one finds
\begin{equation}
     P = \frac{1}{{8 \pi r_+^2}}\left(\frac{Q^2}{r_+^2}-\frac{r_+}{\sqrt{r_+^2-a^2}}-3 c \omega  r_+^{-3 \omega -1}\right)\,,
\end{equation}
which for small $a$ reads
\begin{equation}\label{Pa}
     P = \frac{1}{{8 \pi r_+^2}}\left(\frac{Q^2}{r_+^2}-1 - \frac{a^2}{2 r_+^2}
     -3 c \omega  r_+^{-3 \omega -1}\right)\,.
\end{equation}
This shows that the pressure for zero temperature and  fixed horizon radius increases with $Q$ but decreases with $a$. Since $\omega$ considered here is negative, we also see that the pressure at zero temperature increases with $c$ and with the modulus of $\omega$.

For the complete discussion of the model one has to resort to numerical methods which results will be presented in the next section.


\section{Numerical results: Inversion temperature and Isenthalpic  curves}\label{results}

In this section, we will present our numerical results for the inversion temperature, as well as the isenthalpic  curves. Following the literature \cite{Kastor:2009wy,Kubiznak:2012wp}, we identify the black hole enthalpy with its mass. Then, the isenthalpic curves correspond to  constant black hole masses in the $T-P$ phase diagram.

\begin{figure}[!ht]
\vskip 0.5cm
	\centering
	\includegraphics[scale = 0.40]{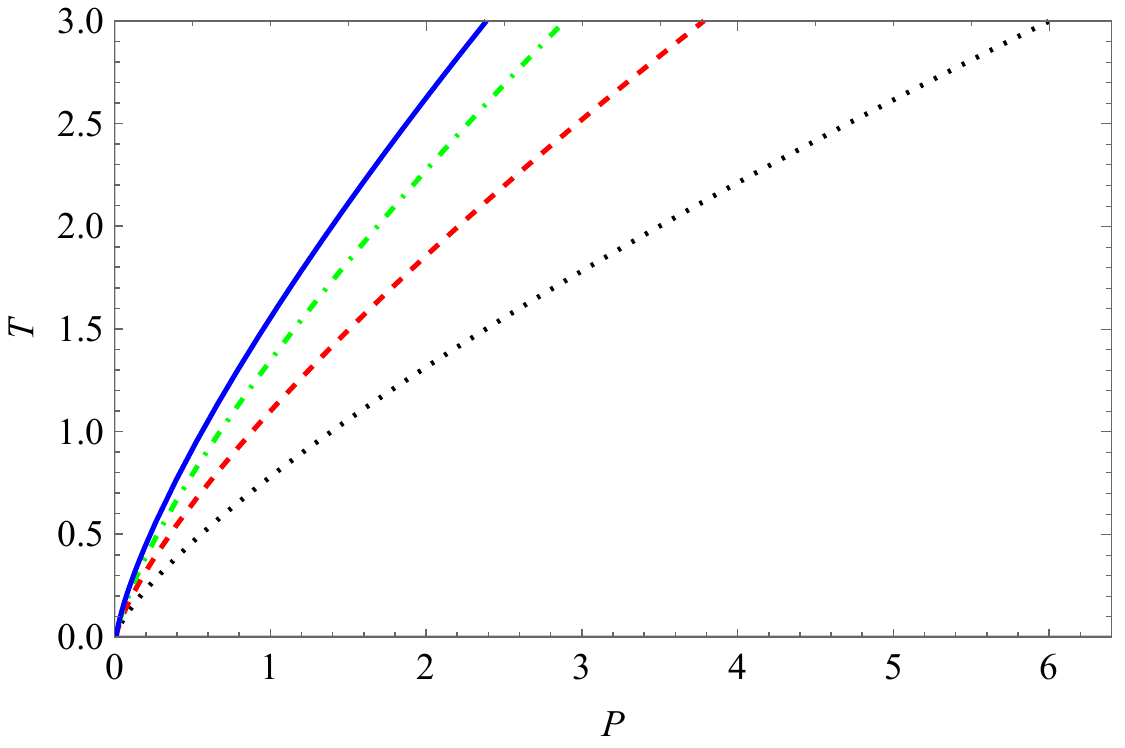}
	\includegraphics[scale = 0.415]{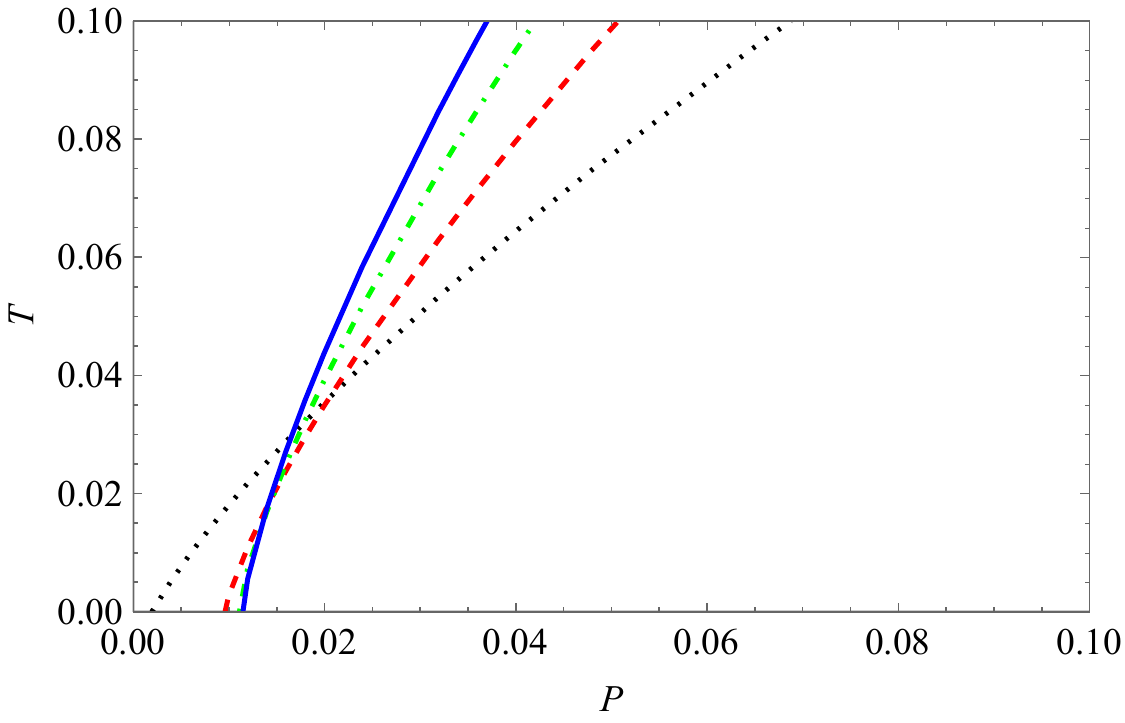}
	\vskip -0.5cm
	\caption{Inversion temperature as a function of the pressure for fixed $\omega = -1.0$, $c= 0.1$,  $a=0.1$ and the following charges $Q=1.0$ (black dotted), $Q=2.0$ (red dashed), $Q=3.0$ (green dot-dashed) and $Q=4.0$ (blue solid). {\sl Left panel:} large range. {\sl Right panel:} small range. We see in these two panels that increasing the electric charge $Q$ increases the  inversion temperature $T$ and its derivative, while the contrary occurs for very small values of temperature and pressure. Note also, in the right panel, the existence of non-zero pressures for $T=0$, and the dislocation of the curves for increasing electric charge $Q$.}
	\label{fig1}
\end{figure}

\begin{figure}[!ht]
\vskip 0.5cm
	\centering
		\includegraphics[scale = 0.40]{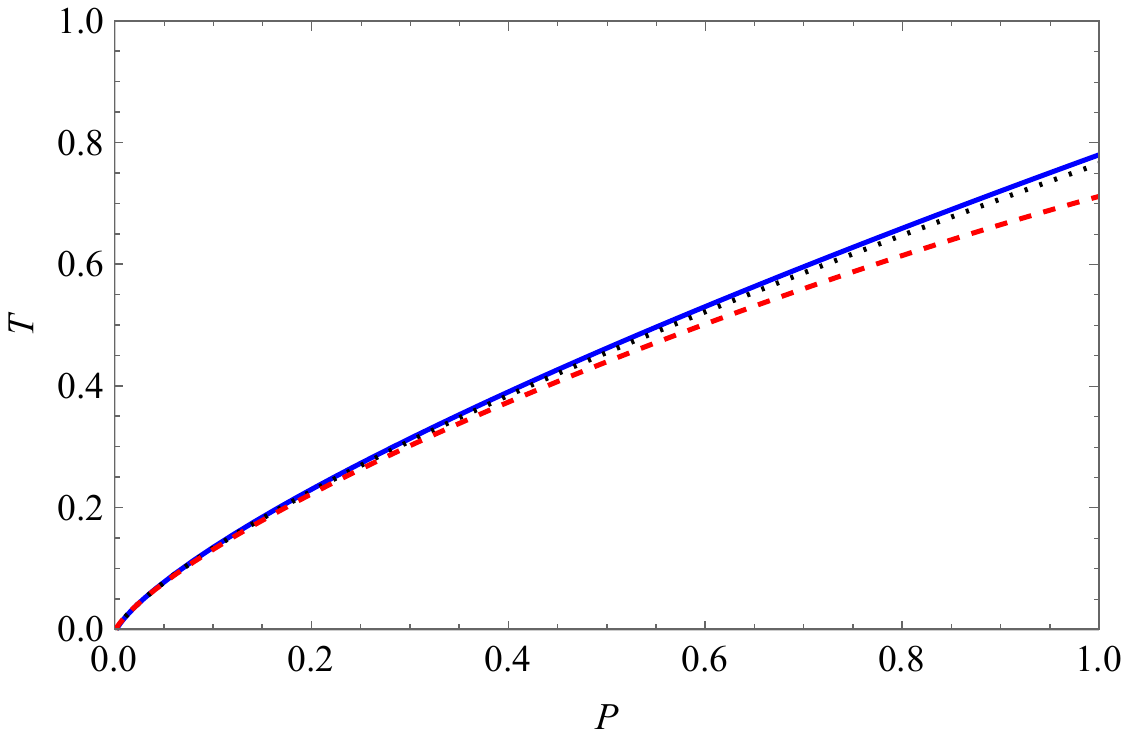}
		\includegraphics[scale = 0.42]{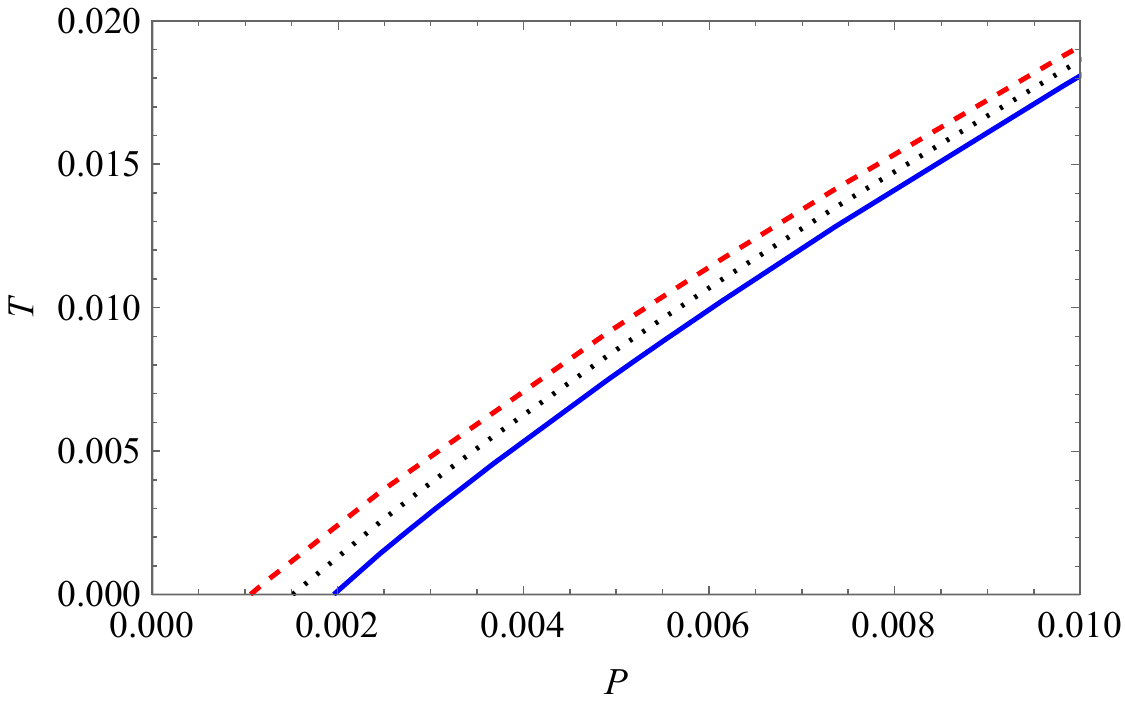}
	\vskip -0.5cm
	\caption{Inversion temperature as a function of the pressure for fixed $\omega = -1.0$, $c= 0.1$, $Q= 1.0$ and varying the quantum correction $a$ with values $0.1$ (solid blue), $0.3$ (dotted black) and $0.4$ (red dashed). {\sl Right panel:} large range. {\sl Left panel:} small range. In this case we see the existence of non-zero pressures for $T=0$, and the dislocation of the curves for {\sl decreasing} quantum correction~$a$. }
	\label{fig2}
\end{figure}

\begin{figure}[ht] 
\vskip 0.5cm
	\centering
	\includegraphics[scale = 0.40]{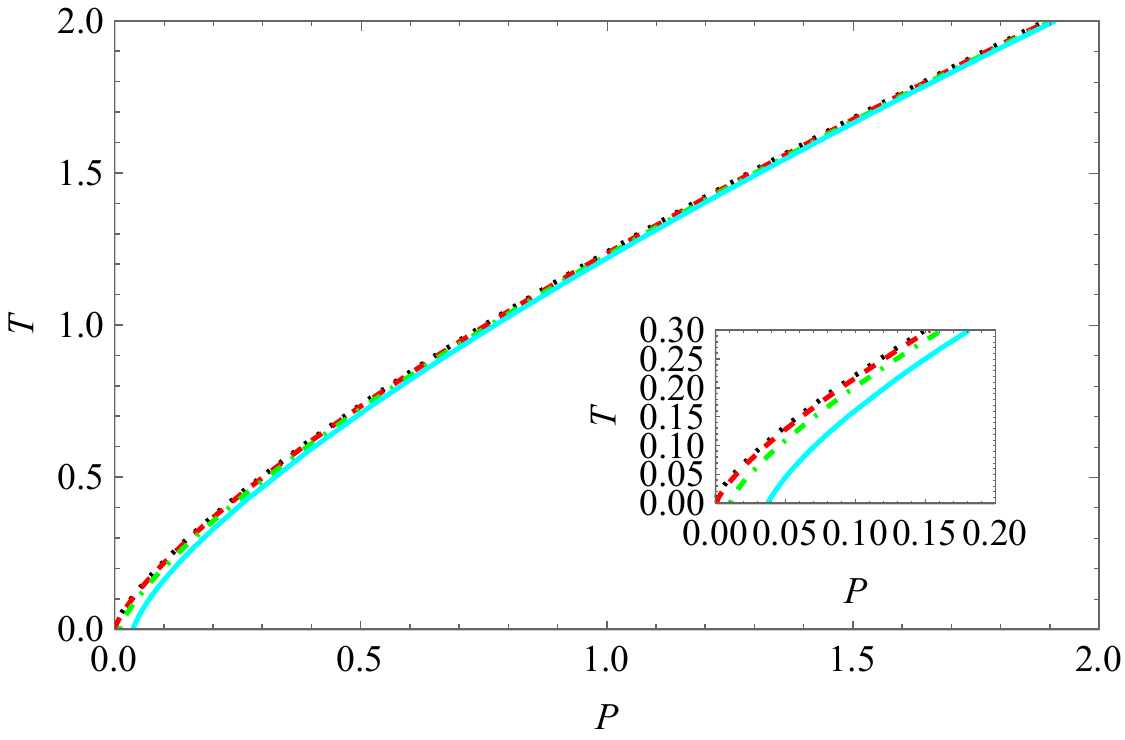}
	\vskip -0.5cm
	\caption{Inversion temperatures as a function of the pressure for fixed charge $Q=2.5$, quantum correction $a= 0.1$ and $c= 0.1$; we vary $\omega$ with the following values and $ -1/6$ (black dotted), $-2/3$ (red dashed), $-1.0$ (green dot-dashed) and $-4/3$ (cyan solid).}
	\label{fig3}
\end{figure}

\begin{figure}[!ht] 
\vskip 0.5cm
	\centering
	\includegraphics[scale = 0.41]{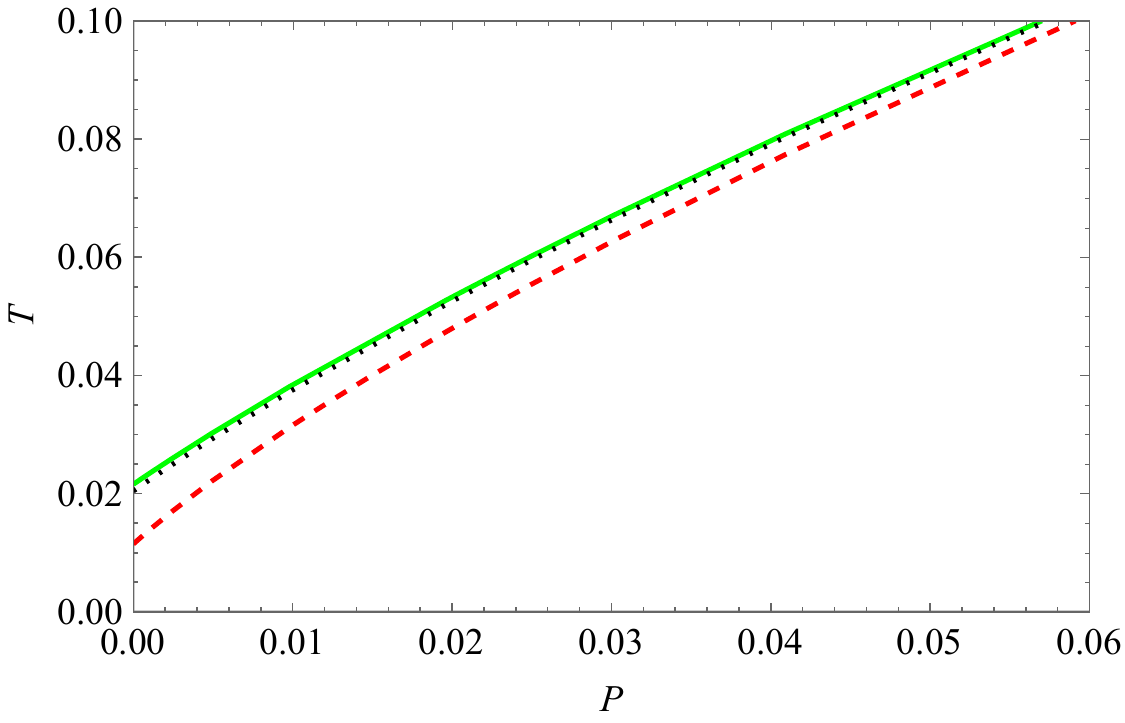}
	\includegraphics[scale = 0.405]{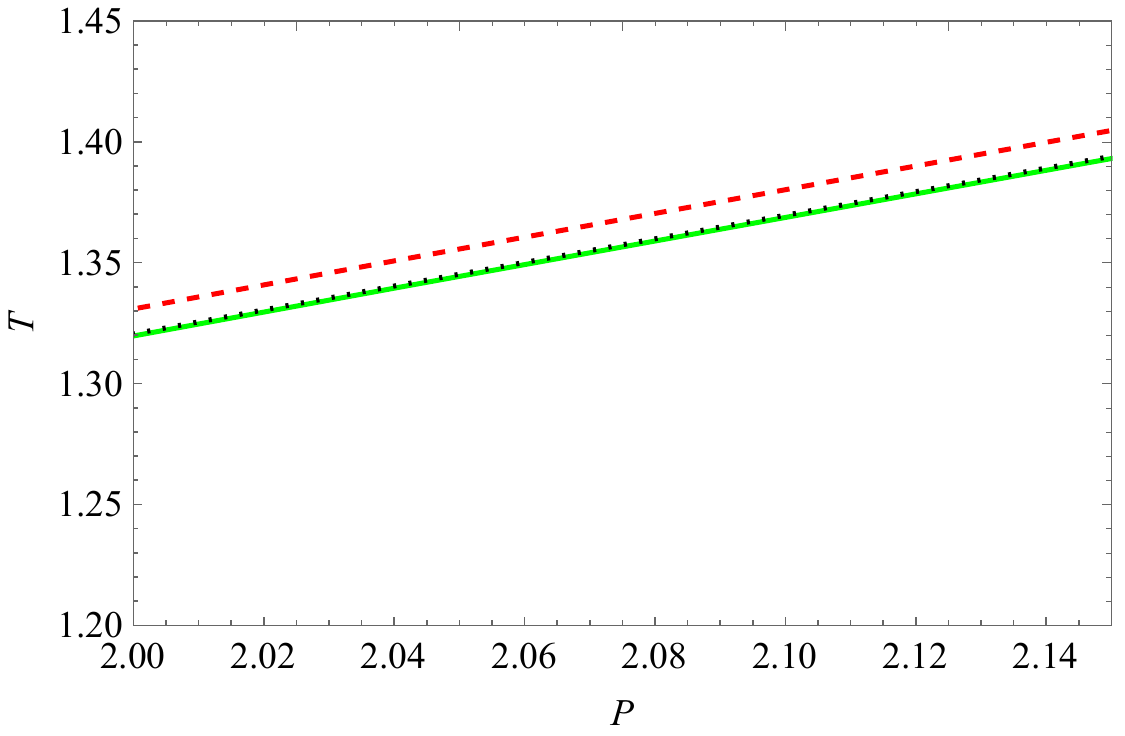}
	\vskip -0.5cm
	\caption{Inversion temperatures as a function of the pressure for fixed charge $Q=1.0$, quantum correction $a= 0.1$, and $\omega = -1/6$. We vary $c$ with the following values $ 0.01$ (green solid), $0.10$ (black dotted), and $1.00$ (red dashed), in different scales ({\sl left and right panels}) to show the details of this function. }
	\label{fig4}
\end{figure}

\begin{figure}[!ht] 
\vskip 0.5cm
	\centering
	\includegraphics[scale = 0.41]{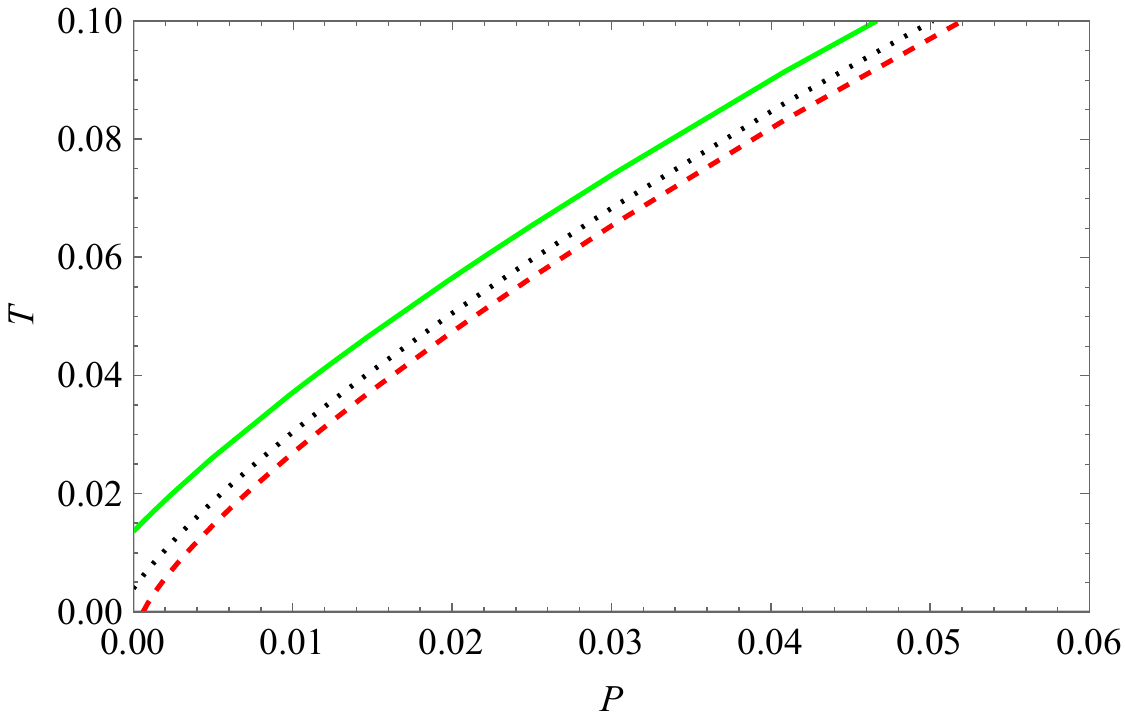}
	\includegraphics[scale = 0.40]{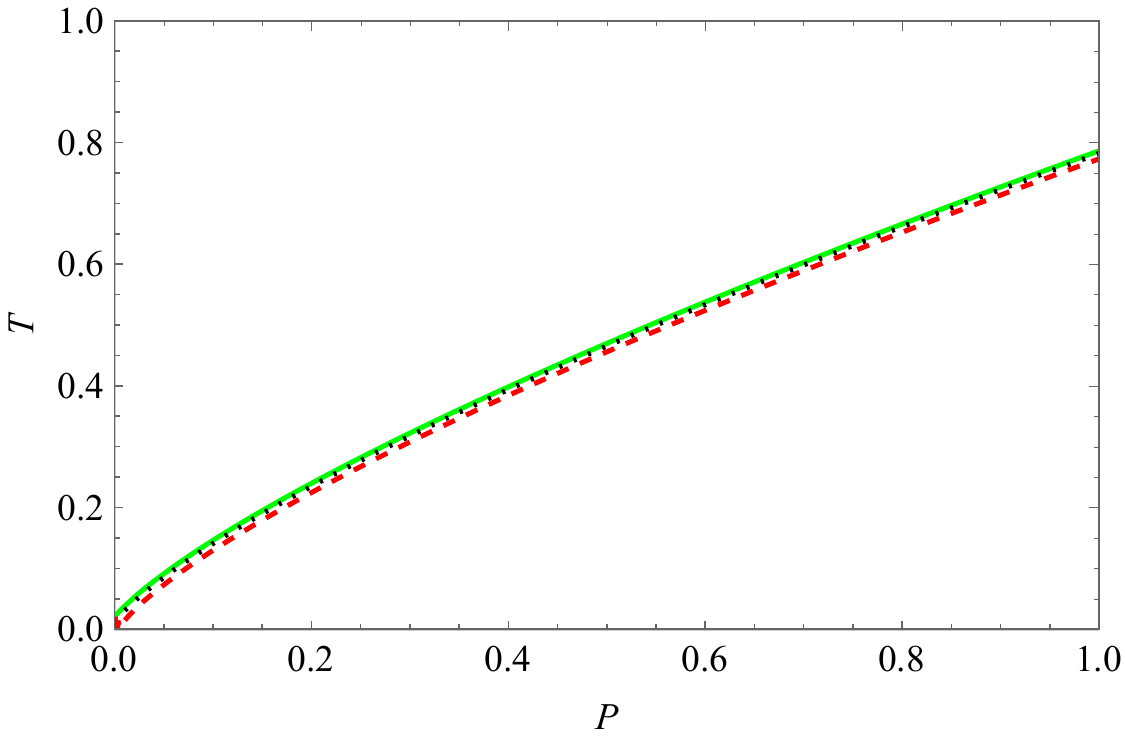}
	\vskip -0.5cm
	\caption{Inversion temperature as a function of the pressure for fixed charge $Q=1.0$ and quantum correction $a= 0.1$. We set $\omega = -2/3$ and vary $c$ with the following values and $ 0.01$ (green solid), $0.10$ (black dotted), and $0.15$ (red dashed).}
	\label{fig5}
\end{figure}

\begin{figure}[!ht] 
\vskip 0.5cm
	\centering
	\includegraphics[scale = 0.40]{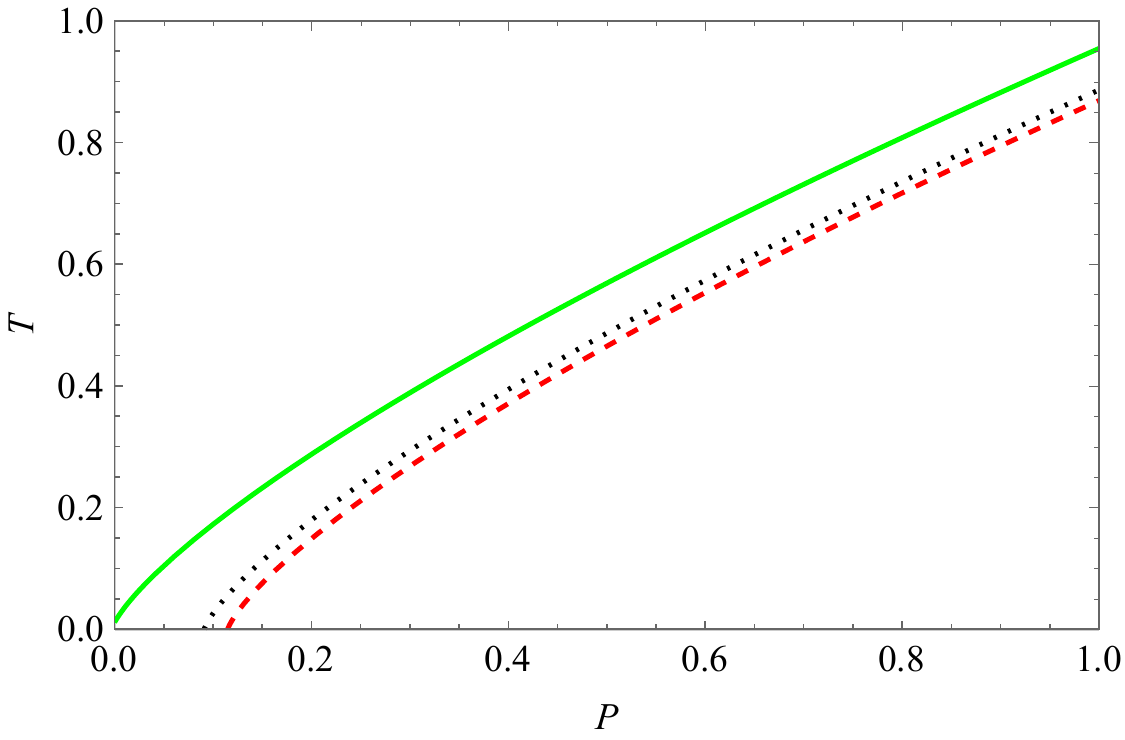}
	\includegraphics[scale = 0.40]{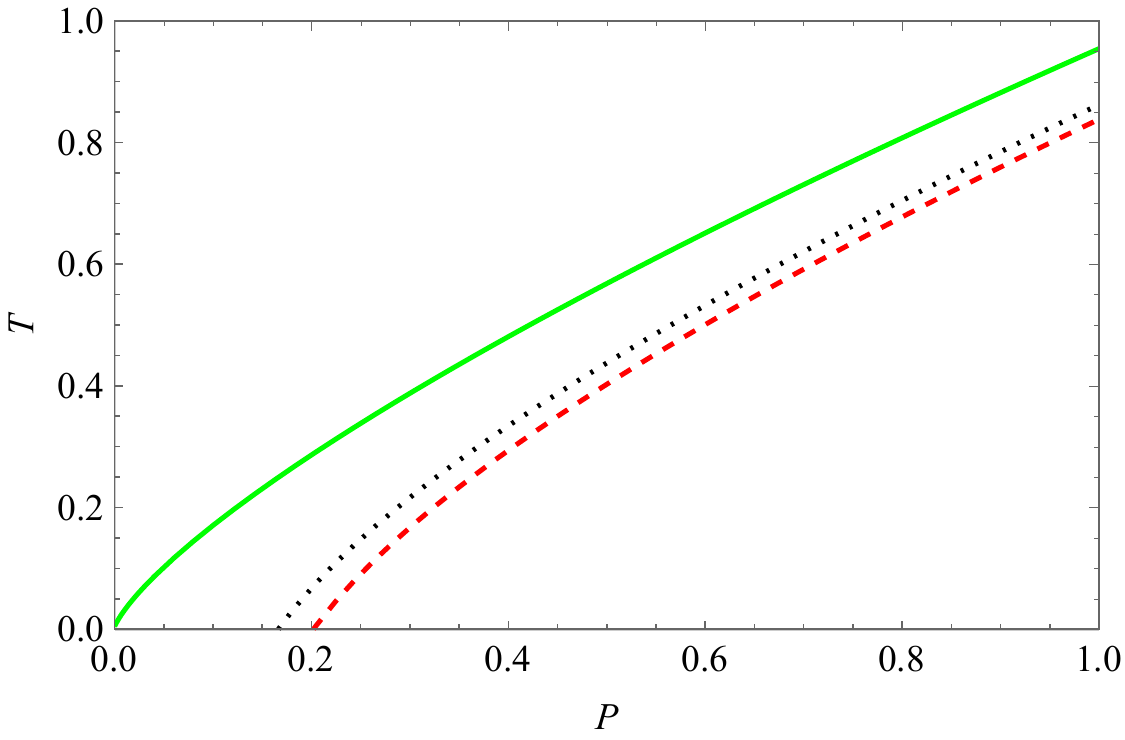}
	\vskip -0.5cm
	\caption{Inversion temperatures as a function of the pressure for fixed charge $Q=1.0$ and quantum correction $a= 0.1$. {\sl Left panel:}  we set $\omega = -1$ and vary $c$ with the following values and $ 0.01$ (green solid), $0.80$ (black dotted), and $1.00$ (red dashed).  {\sl Right panel:} we set  $\omega = -4/3$ and vary $c$ with the following values and $ 0.01$ (green solid), $0.80$ (black dotted), and $1.00$ (red dashed).}
	\label{fig6}
\end{figure}

First, note that the Eq. \eqref{dift} for $r_+$ does not have analytical solutions for arbitrary values of the parameters $a$, $c$, $\omega$, and $Q$. So, we solve it numerically, obtaining $r_+=r_+(a,c,\omega,Q,P)$. Then, we  use these solutions in Eq. \eqref{ti1} to obtain the inversion temperature $T=T(a,c,\omega,Q,P)$. These results are presented in Figs. \ref{fig1} - \ref{fig6},  where we plot  $T$ as a function of the pressure $P$, for some choices of the parameters $a$, $c$, $\omega$, and $Q$.

In the left and right panels  of Fig.~\ref{fig1}, we have fixed the values of $\omega=-1$, $c=0.1$, $a=0.1$ and varied  the electric charges from $Q=1.0$ to $Q=4.0$.  
In left panel of Fig.~\ref{fig1}, we see as expected from the literature, for instance  \cite{Okcu:2016tgt,Okcu:2017qgo,Mo:2018rgq,Lan:2018nnp}, that raising the black hole electric charge $Q$ the inversion temperature $T$ increases. However, the opposite happens for very small values of the temperatures and pressures, as can be seen in the Right panel of Fig. \ref{fig1}.

In the right panel of Fig.~\ref{fig1}, we see that there are non-zero minimum values for the pressure $P$ at $T=0$. In general, for an AdS-RN black hole, there is a minimum temperature and not a minimum pressure; however, in our model, for non-zero values of the parameter $c$, the opposite occurs: the model no longer has a minimum temperature, but a minimum pressure. This behavior seems to be a feature of models with some kind of cosmological fluid, as seen in \cite{K.:2020rzl}, where the cases $\omega = -1$ and $\omega = -1/3$ has been discussed for a Bardeen black hole.

In the right panel of Fig.~\ref{fig1}, we also see that these non-zero values for the pressure increase with increasing electric charge $Q$. So, besides the increase of the derivative of the inversion temperature, the increase of the electric charge $Q$ also implies a small displacement of these curves in the direction of increasing pressure, in accordance with Eq. \eqref{Pa}. In this panel, one also sees that for small values of the temperature and pressure the curves of constant charge intersect each other.


%
In the left and right panels  of Fig.~\ref{fig2}, we plot the inversion temperature against the pressure for fixed values of $\omega=-1$, $c=0.1$, $Q=1.0$ and varying the quantum corrections from $a=0.1$ to $a=0.4$. 
In the left panel of Fig.  \ref{fig2},
we see that the increase of the quantum correction decreases the inversion temperature. One should note that this happens for the high pressure regime. 
In the right panel of Fig.  \ref{fig2},  we see that there are non-zero minimum values for the pressure $P$ at $T=0$, and that these non-zero values for the pressure decrease with increasing quantum correction $a$. So, besides the increase of the derivative of the inversion temperature, the increase of the quantum correction also implies a small dislocation of these curves in the direction of decreasing pressure, in agreement with Eq.~\eqref{Pa}. 

Comparing the left and right panels of  Fig.  \ref{fig2}, we conclude that these curves should intercept each other at intermediate values of the pressure.

In Fig. \ref{fig3}, we present the inversion temperature versus the pressure for fixed $Q=2.5$, $a=0.1$, $c=0.1$, and  varying the quintessence $\omega=-1/6$, -2/3, -1, and -4/3.  
For increasing the absolute value of the quintessence we see a diminishing of the inversion temperature, or equivalently the increase of the pressure, as expected from Eq. \eqref{Pa}. 
In particular, it seems that for the phantom dark energy case  $(\omega=-4/3)$ there might be a minimum inversion pressure. 


In Fig. \ref{fig4}, we present the inversion temperature for fixed black hole charge, quantum correction,  $\omega=-1/6$, and vary the cosmological fluid $c$ from 0.01 to 1.00. In the left panel, when we increase $c$ the pressure increases, consonantly with Eq. \eqref{Pa}, and the inversion temperature decreases. 
We also see a minimum inversion temperature for each value of $c$. In particular, when we increase $c$, we see that the minimum temperature decreases. 
In the right panel, the opposite occurs, which means, when we increase $c$ the pressure diminishes and the temperature increases. Comparing the two panels, we see that these curves should intercept each other for some intermediate pressure.


In Fig. \ref{fig5}, we show the inversion temperature for fixed black hole charge, quantum correction,  $\omega=-2/3$, and vary the cosmological fluid $c$ from 0.01 to 0.15. In the left panel, when we increase $c$ the pressure increases as predicted by Eq. \eqref{Pa} and the inversion temperature decreases.  
We see minimum values for the inversion temperature only for $c$=0.01 and 0.10. In particular, for $c=0.15$ there is no minimum temperature but there is a minimum pressure. 
In the right panel, when we increase $c$ the pressure increases and the temperature decreases, as in the left panel. In this case we do not see the interception of the curves.


%

Now, we plot in Fig. \ref{fig6}, the inversion temperature for fixed quantum correction and black hole charge, and span the cosmological fluid $c$ from 0.01 to 1.0. In the left panel, we choose $\omega=-1$, while in the right panel, we took $\omega=-4/3$. In both cases, if we increase $c$ the pressure increases as expected from Eq. \eqref{Pa} and the inversion temperature decreases. 
We also see a minimum value for the inversion temperature only for $c$=0.01. On the other hand, for $c=0.80$ and 1.0 there are  minimum values for the pressure. 
 Notice that we do not see the interception of the curves for fixed $\omega$ and  comparing the left and right panels one sees that if we increase the absolute value of $\omega$ the minimum pressure increases as well, also in agreement with Eq. \eqref{Pa}.


\begin{figure}[ht] 
\vskip 0.5cm
	\centering
	\includegraphics[scale = 0.40]{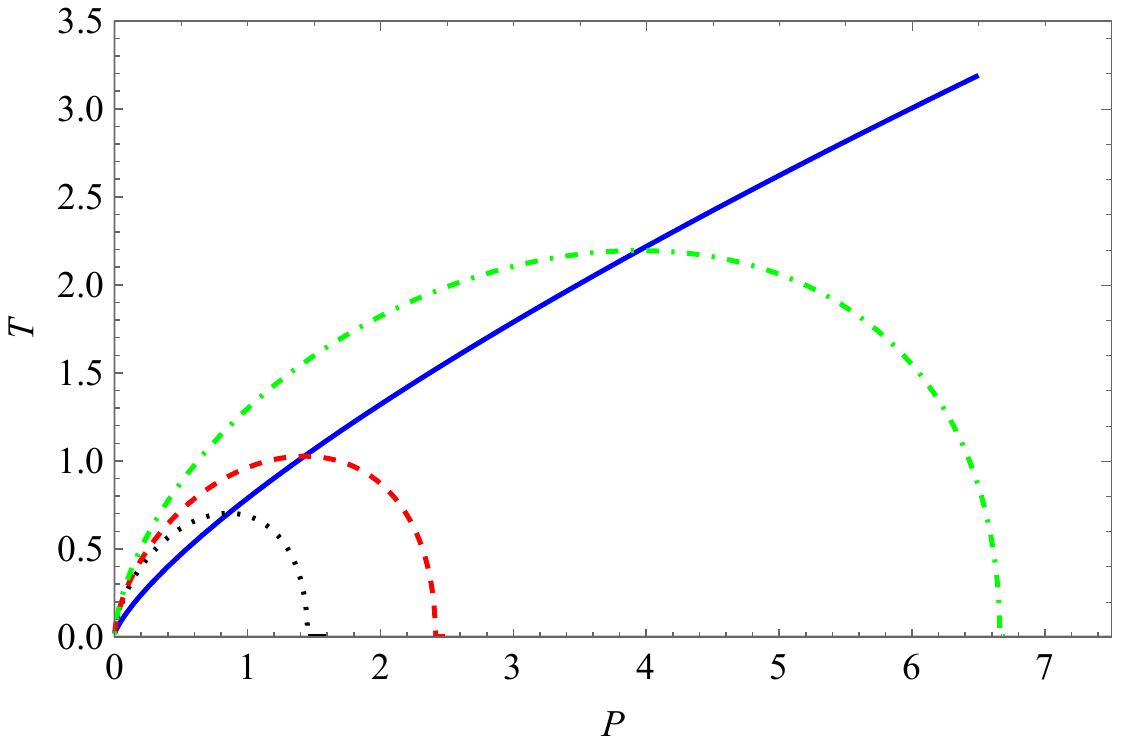}
	\includegraphics[scale = 0.40]{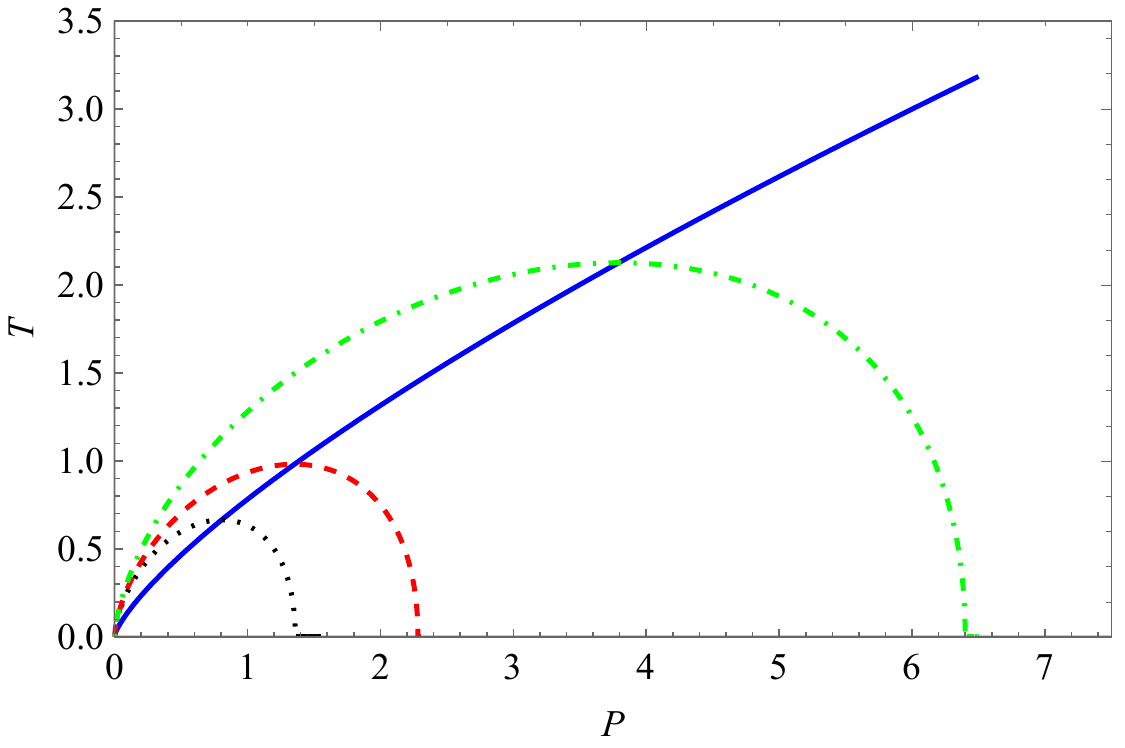}
	\includegraphics[scale = 0.40]{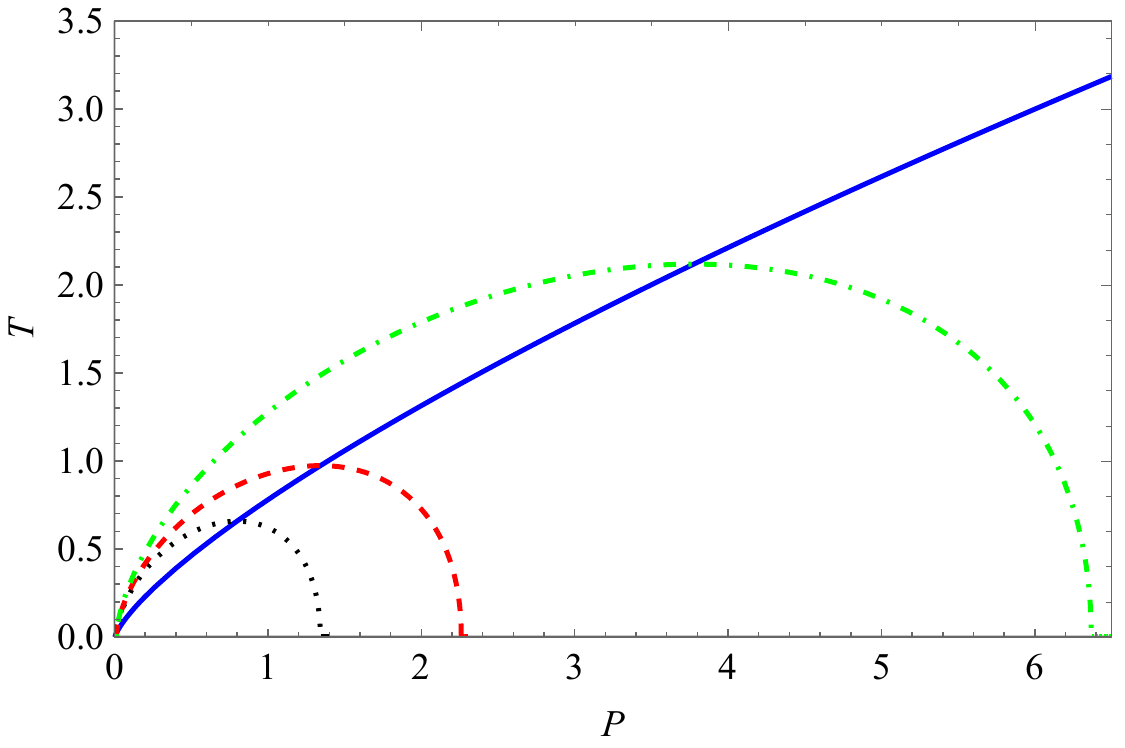}
	\includegraphics[scale = 0.40]{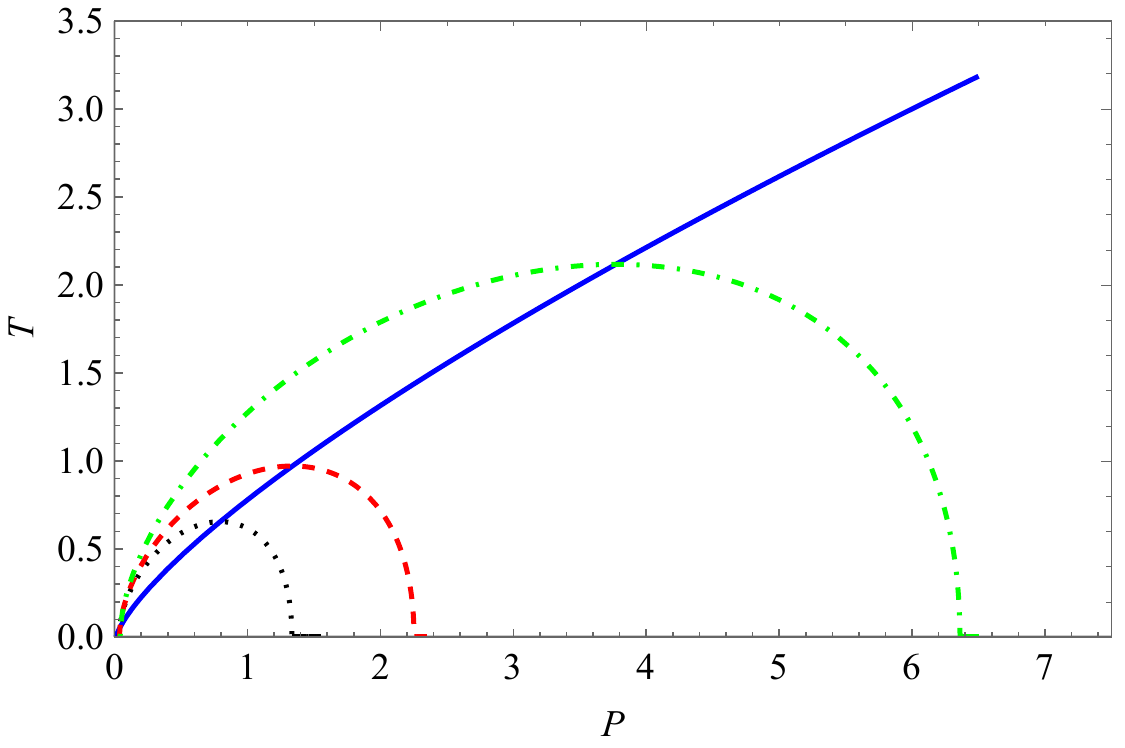}
	\vskip -0.5cm
	\caption{Isenthalpic curves for different values of the black hole mass, $M=1.8$ (black dotted), $M=2.0$ (red dashed), $M=3.0$ (green dot-dashed), for fixed $a= 0.1,\,Q= 1.0$, and  $c= 0.1$. The blue solid curves represent the inversion temperatures.   {\sl Upper left panel:} $\omega = -1/6$. {\sl Upper right panel:}  $\omega=-2/3$. {\sl Lower left panel:} $\omega = -1$. {\sl Lower right panel:} $\omega = -4/3$.}
	\label{fig7}
\end{figure}

In order to obtain the isenthalpic curves for the black hole, we solve numerically the Eq. \eqref{mbh} for $r_+$ to get 
 $r_+=r_+(a,c,\omega,P, Q, M)$, where we used the relation between the pressure and the AdS radius, Eq. \eqref{pressao}.  Applying this solution in Eq. \eqref{ti1} we achieve the profiles of the black hole mass in a plane $T-P$ for fixed sets of the other parameters. These results are shown in Fig. \ref{fig7}.  
  Note that the isenthalpic curves are very similar for fixed values of $Q$, $a$ and $c$,   varying $\omega$.
 However, there are small differences: in particular the values where the isenthalpic curves touch the pressure axis decrease as the absolute value of $\omega$ increases. Note that, as expected, the inversion temperature curves intercept the insenthalpics at their maxima. 
 This  can be understood as a numerical check of our analysis.

\section{conclusions}\label{conc}

In this paper, we study the Joule-Thomson expansion  for an AdS-RN black  hole  taking into account quantum corrections, in the scenario of a Kiselev metric. Within our model, we also considered the influence of a cosmic fluid. In order to mimic such a fluid, we have used equations of states given by $P = \omega \rho $, with $\omega$ assuming different values. 
For $\omega=-1$ one has the cosmological constant, for $\omega=-1/6$ or $-2/3$ one has quintessence, and phantom dark energy with $\omega =-4/3$. For each model, our goal was to study the behavior of the inversion temperature as a function of the pressure.

At this point it is worthwhile to make a first comment on the  analysis of our results.  If we consider only quantum corrections, without cosmological fluid $(c=0)$ or with a cosmological fluid that can be neglected $(c\to 0)$, we are able to perform a semi-analytical study and explicitly find an equation for the minimum inversion temperature as a function of the quantum correction parameter, at least in first order in $a$, as can be seen in Eq. \eqref{tmin}. This result shows us that the model has a non-null minimum inversion temperature in this case ($c=0$ or  $c\to 0$) . 

When we considered quantum corrections and the cosmological fluid, or just the cosmological fluid, we were unable to find  semi-analytical results for the inverson temperature, so we did a numerical study. In this case, we find that the introduction of the cosmological fluid not only changes the shape of the minimum temperature inversion curves, but can also lead to a minimum inversion pressure, as described by Eq. \eqref{Pa}. This feature seems to be typical of models with cosmological fluids, considering that it was also found in \cite{K.:2020rzl}, in the case where $\omega = -1/3 $ for an AdS-RN Bardeen black hole. It is interesting to note how cosmological fluids can bring the minimum inversion temperature to zero.

In all the studied cases, the slope of the temperature inversion curves are strictly positive, however their shape varies with the values of the model parameters, for instance, the values for the quantum correction $a$ and the cosmological fluid $c$.  In order to verify the consistency of our study, we also plotted the isenthalpic curves, showing that the temperature inversion curves intersect the isenthalpics at their maxima as expected. 

We hope that this study might help understand black hole thermodynamics in different scenarios. 

\begin{acknowledgments}

The authors would like to thank an anonymous Referee for criticisms and suggestions which help improve the quality of this work, Danning Li for numerical support and Shihao Bi for useful conversations. J.P.M.G is supported by Conselho Nacional de Desenvolvimento Científico e Tecnológico (CNPq)  under Grant No. 151701/2020-2.  H.B.-F. is partially supported by  Conselho Nacional de Desenvolvimento Científico e Tecnológico (CNPq) under Grant No. 311079/2019-9. This work was also supported by  Coordenação de Aperfeiçoamento de Pessoal de Nível Superior (CAPES) under finance code 001. I. P. L. was partially supported by the National Council for Scientific and Technological Development - CNPq grant 306414/2020-1 and by the grant 3197/2021, Para\'iba State Research Foundation (FAPESQ). I. P. L. would like to acknowledge the contribution of the COST Action CA18108.
 
\end{acknowledgments}


\end{document}